\documentclass{PoS}
\usepackage{graphicx}

\PoS{PoS(LAT2005)123}

\title{Overlap fermion with the topology conserving gauge action}

\ShortTitle{Overlap fermion with the topology conserving gauge action}

\author{Hidenori~Fukaya$^a$, Shoji~Hashimoto$^{b,c}$,
  Takuya~Hirohashi$^{d}$, 
  \speaker{Hideo~Matsufuru}$^{b}$
  Kenji~Ogawa$^{c}$, and Tetsuya~Onogi$^{a}$\\
  \llap{$^a$}Yukawa Institute for Theoretical Physics, Kyoto University,
             Kyoto 606-8502, Japan.\\
  \llap{$^b$}High Energy Accelerator Research Organization (KEK),
             Tsukuba 305-0801, Japan.\\
  \llap{$^c$}School of High Energy Accelerator Science,
             The Graduate University for Advanced Studies (Sokendai),
             Tsukuba 305-0801, Japan.\\
  \llap{$^d$}Department of Physics, Kyoto University,
             Kyoto 606-8502, Japan.\\
  E-mail: \email{hideo.matsufuru@kek.jp}
}

\abstract{
  We investigate the distribution of low-lying eigenmodes
  of hermitian Wilson-Dirac operator, $H_W$, with the gauge
  action whose form is designed to avoid topology change,
  as well as with the standard plaquette action.
  On the quenched lattices, the former gauge action
  exhibits less density of low-lying eigenmodes of $H_W$
  compared to the latter at the same lattice spacing.
  We also show preliminary results for the dynamical
  simulation with two flavors of overlap fermions. 
}

\FullConference{XXIIIrd International Symposium on Lattice Field Theory\\
                 25-30 July 2005\\
                 Trinity College, Dublin, Ireland}

\begin{document}

\section{ Introduction }

Recent progress of the chiral fermions,
both in theoretical understanding and in technical
development of numerical algorithms, is making its dynamical
fermion simulation more feasible with the next generation
machines. 
Wide range of applications from flavor physics to hadron
physics is expected with the dynamical fermions having exact
chiral symmetry.

The Neuberger's overlap-Dirac operator with mass $m$ is
written as \cite{Neuberger:1997fp}
\begin{equation}
  \label{eq:ov}
  D_{ov}(m) = \left(1 - \frac{\bar{a}}{2}\right) D_{ov} + m,
  \;\;\;
  D_{ov} = \frac{1}{\bar{a}} [1 + \gamma_5 \mbox{sign}(H_W) ],
\end{equation}
where $\bar{a}\equiv a/(1+s)$.
For the kernel $H_W$ we use the Wilson-Dirac operator $D_W$ as
\begin{equation}
  H_W = \gamma_5 [ D_W - (1+s) ].
\end{equation}
The parameter $s$ is chosen in the range $|s|<1$.
The locality of the overlap operator is proven when the
bulk of eigenvalues of $H_W$ is bounded in a non-zero region 
\cite{Hernandez:1998et}.
This condition is satisfied when the gauge configuration is
sufficiently smooth.
In practice, however, it is known that there is non-zero
density of near-zero eigenvalues when the gauge
configurations are generated with the conventional Wilson
gauge action \cite{Edwards:1998sh}.
Although the problem of locality is avoided if the near-zero
modes are localized
\cite{Golterman:2003qe,Golterman:2004cy,Golterman:2005fe}, 
it is more desirable if the gauge action does not generate
such near-zero modes from the beginning.
The use of improved actions (L\"uscher-Weisz, Iwasaki, DBW2)
is a popular choice, with which the density of the
near-zero modes is reduced,
but in this work we consider an alternative solution.

Since the zero mode of $H_W$ appears when the topological 
charge of the gauge field changes its value, we expect that
the appearance of the near-zero mode is suppressed if we use
a gauge action for which the topology change is suppressed
or even forbidden.
In this work, we employ the L\"uscher's topology conserving
gauge action \cite{Luscher:1998du} and investigate how much
it suppresses the near-zero modes.

The suppression of topology change and of the appearance of
the near-zero modes is desirable also in a practical point
of view. 
(1) The cost of approximating the sign function in
(\ref{eq:ov}) is reduced;
(2) In the dynamical fermion simulation, one can avoid the
``reflection-refraction'' process \cite{Fodor:2003bh}, which
is very costly;
(3) Gauge configurations with a fixed topological charge is
efficiently accumulated, which is useful for the study of
the $\epsilon$-regime of chiral perturbation theory.

\section{Topology conserving gauge action}
\label{sect:gauge_action}

The topology conserving gauge action can be constructed by
enforcing the admissibility condition
\begin{equation}
  \label{eq:addmissibility}
  | 1 - P_{\mu\nu}(x) | < \epsilon, \hspace{1cm}
  \mbox{$\forall x,\mu,\nu$},
\end{equation}
where $P_{\mu\nu}(x)$ is the plaquette.
$\epsilon$ is a parameter to control the smoothness of the
gauge field.
For the SU(3) gauge theory, the locality of the
overlap-Dirac operator is guaranteed when $\epsilon\lesssim$ 
1/20.49 \cite{Neuberger:1999pz}.
In practice this condition is too strong to realize the
lattice spacing $\sim$ 0.1~fm and we have to relax it to
$\epsilon\sim O(1)$.
Therefore, the question is whether the good property is kept
enough for these large value of $\epsilon$.

An example of the gauge action to enforce
(\ref{eq:addmissibility}) is 
\begin{equation}
 S_G = \left\{ 
   \begin{array}{ll} \displaystyle
          \frac{1}{g^2} \sum_{x,\mu,\nu} \frac{S_P(x)}
                          {(1-S_P(x)/\epsilon)^\alpha}
          & \hspace{1cm} \mbox{if admissible} \\ 
   \infty & \hspace{1cm} \mbox{otherwise} \\
 \end{array}
 \right.
\end{equation}
where $S_P(x) = 1-\mbox{ReTr}P_{\mu\nu}(x)/N_c$.
In this work, we set $\alpha =1$.
Previously, this gauge action was used by two of us for the
study of the massive Schwinger model \cite{Fukaya:2003ph}.
In the four-dimensional QCD, Bietenholz {\it et al.} are
also studying its property
\cite{Shcheredin:2004xa,Bietenholz:2004mq}.

In the following, we investigate the eigenvalue distribution
of $H_W$ on the gauge configurations generated with this
action.
The topology change and scaling property of this action is 
presented in these proceedings \cite{Onogi}.

\section{Quenched results}
\label{sect:quenched}

First we examine the low-lying eigenvalue distribution of $H_W$
for the topology conserving gauge action in the quenched
approximation.
The configuration is generated with the Hybrid Monte Carlo
algorithm with $\delta t=$ 0.01 and a length of trajectory 0.2.
The parameter $\epsilon$ is set to $1/\epsilon=$ 
0, 2/3, and 1.
($1/\epsilon=$0 corresponds to the standard plaquette action.)
The admissibility condition, $1-S_P(x)/\epsilon >0$, is
monitored during the molecular dynamics evolutions, and no
violation has been found for $1/\epsilon=$2/3 and 1.
For finite $1/\epsilon$, while the topology change is not
exactly prohibited, it is indeed much suppressed compared to
the plaquette action $1/\epsilon=0$ \cite{Onogi}.

\begin{figure}[tb]
\center{
\includegraphics[width=7cm,clip]{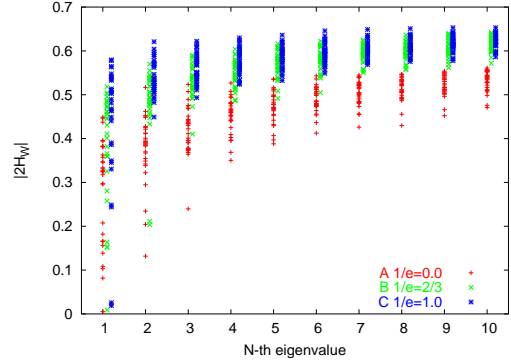}}
\vspace{-0mm}
\caption{
  Low-lying eigenvalue distribution on quenched $16^4$
  lattices for $1/\epsilon=0$, 2/3, and 1 
  ($a^{-1}\simeq 2.5$ GeV).
}
\label{fig1}
\vspace{-0mm}
\end{figure}

\begin{figure}[tb]
\center{
\includegraphics[angle=-90,width=7cm,clip]{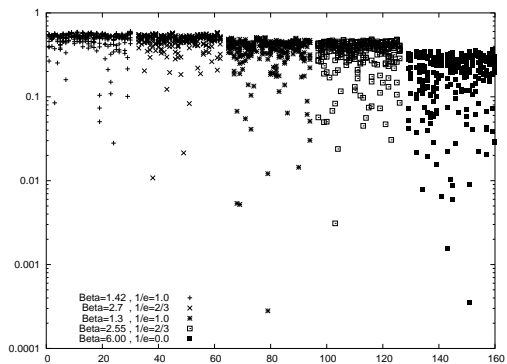}}
\vspace{-0mm}
\caption{
  Low-lying eigenvalue distribution of $2H_W$ on quenched $20^4$ lattices.} 
\label{fig2}
\vspace{-0mm}
\end{figure}

Figure~\ref{fig1} shows ten lowest-lying eigenvalues of $|H_W|$
on a $16^4$ lattice for the three values of $\epsilon$.
The inverse lattice spacing determined with the Sommer scale is 
$a^{-1}\simeq 2.5$ GeV for each lattice.
In the plot, A, B, and C correspond to the combinations
$(\beta,1/\epsilon) = (6.13,0)$, 
$(2.70,2/3)$, and $(1.42, 1)$, respectively.
As expected, we find that the eigenvalue distribution shifts
upward as $1/\epsilon$ increases.

Figure~\ref{fig2} shows the results on $20^4$ lattices.
In this figure, the low-lying eigenvalue distribution is
compared at five sets of parameters:
$(\beta, 1/\epsilon)$ = (6.0,0), (2.55,2/3), (1.3,1),
(2.7,2/3), and (1.42,1).
The first three correspond to $a^{-1}\simeq$ 2.1~GeV,
while the last two are 2.6~GeV and 2.5~GeV.
The same tendency as in Figure~\ref{fig1} is found.

We conclude that in the quenched calculation
the density of the near-zero eigenvalues is smaller for the
topology conserving gauge action than that of the plaquette
action.
This trend is enhanced as $1/\epsilon$ increases.

\section{Dynamical results}
\label{sect:dynamical}

In this section, we show the results of an exploratory study 
of the low-lying eigenvalue distribution of $H_W$
in the dynamical fermion simulation with two flavors of
the overlap fermion. 
Our algorithm of the Hybrid Monte Carl update follows that
of \cite{Fodor:2003bh}.
Namely, we adopt the Zolotarev rational approximation
\cite{Eshof02}
with degree $N_p=20$ to approximate the sign function.
The low-lying eigenmodes of $H_W$ is calculated using the
implicitly restarted Lanczos method.
When the lowest eigenmode changes its sign, 
the value of the pseudofermion action changes
discontinuously. 
Then the reflection or refraction occurs depending on the
energy of the momentum of the mode parpendicular to the
surface corresponding to that eigenmode.
The occurrence of the reflection and refraction is monitored
during the HMC update.

Numerical simulation is performed on $4^4$ lattices,
with overlap fermions at $m=0.2$ and $s = 0.6$.
Configurations are generated with $\delta t=0.02$ and an
unit length of trajectory.
The values of $\beta$ and $1/\epsilon$ are listed in
Table~\ref{tab1}.
Since the lattice spacing is unknown at this stage
and the lattice volume is so small, we cannot draw any
definite conclusion, but we would report the status of our
study.

The average number of occurrence of reflection is listed in
Table~\ref{tab1},
and the refraction is used to trace the change of topological
charge, as shown below.
If the topology conserving action properly works,
appearance of the near-zero mode should be suppressed, and
hence the reflection/refraction should also be suppressed.
Our result does not show clear tendency supporting this
expectation.

\begin{table}[tb]
\begin{center}
\begin{tabular}{ccccc}
\hline\hline
size & $\beta$ & $1/\epsilon$ &  reflection/trj & acceptance \\
\hline
$4^4$ & 5.40 & 0 &  0.34 & 0.97  \\
      & 5.50 & 0 &  0.53 & 0.94  \\
\hline
$4^4$ & 0.70 & 1 &  0.63 & 0.82  \\
      & 0.80 & 1 &  0.18 & 0.89  \\
\hline\hline
\end{tabular}
\end{center}
\vspace{-0.4cm}
\caption{
  Parameters of our dynamical simulation runs.
}
\label{tab1}
\end{table}

\begin{figure}[tb]
\center{
\hspace*{-0.5cm}
\includegraphics[width=7.cm]{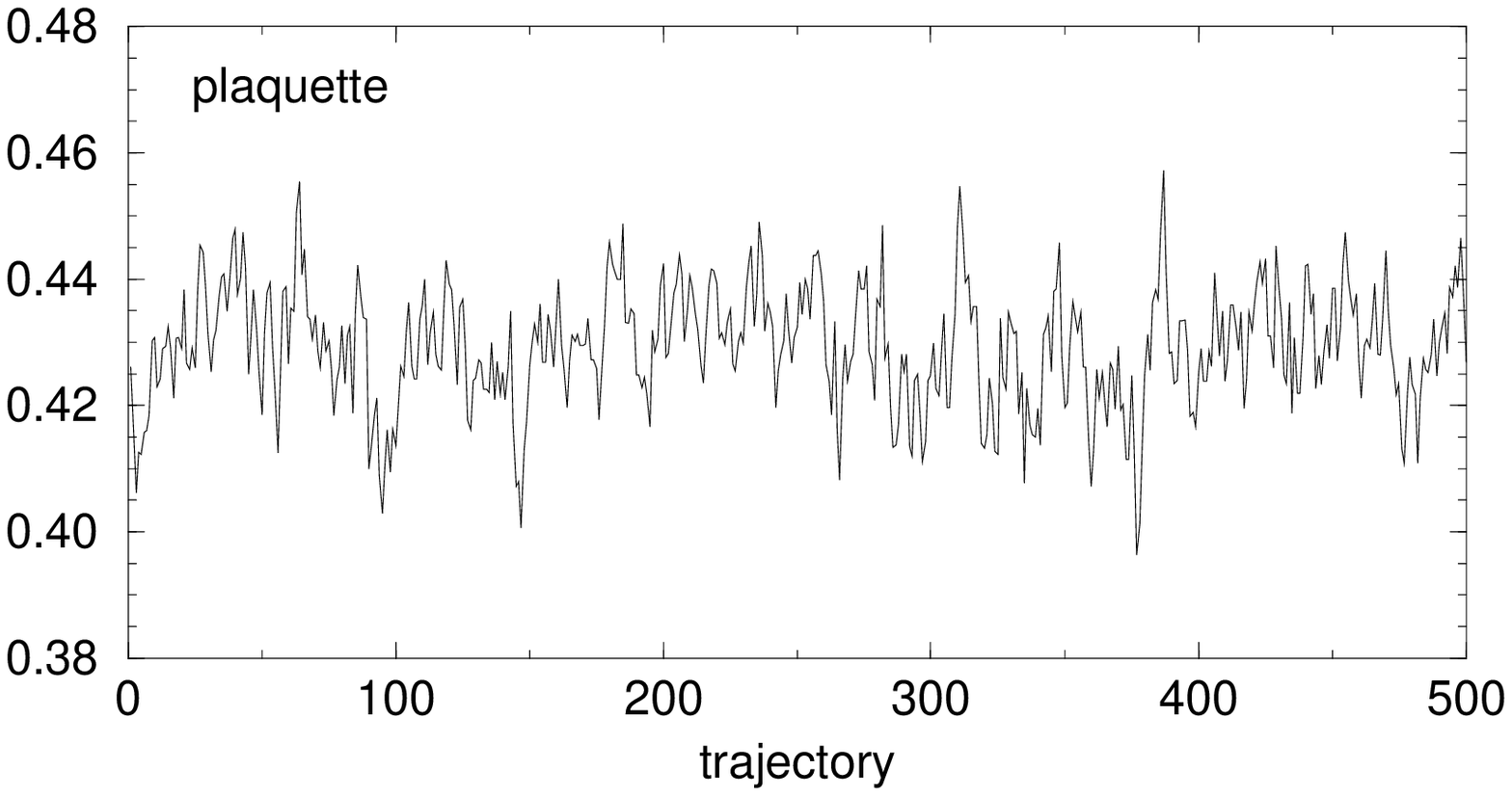}
\includegraphics[width=7.cm]{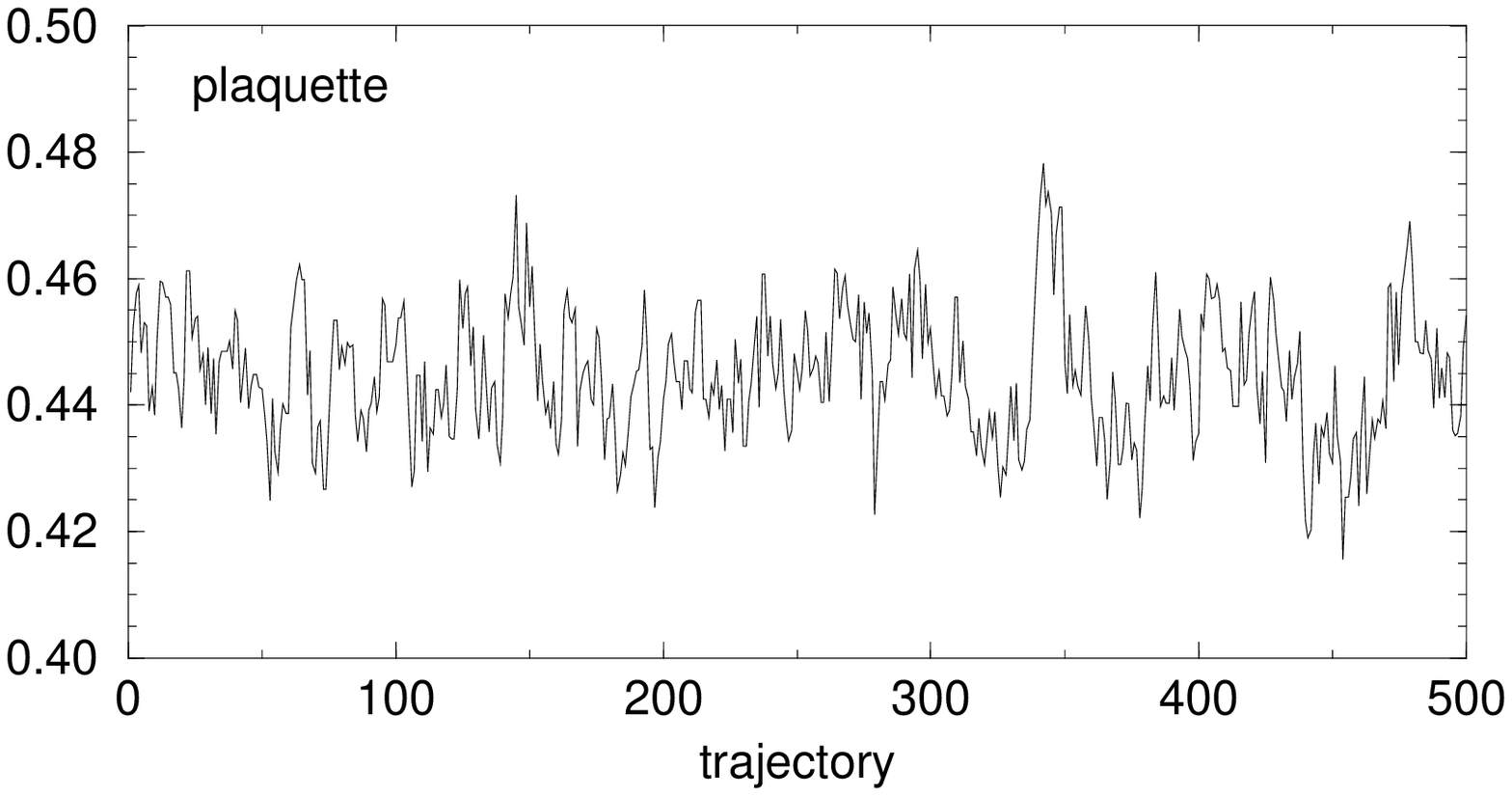}
\vspace{-0.15cm}\\
\hspace*{-0.5cm}
\includegraphics[width=7.cm]{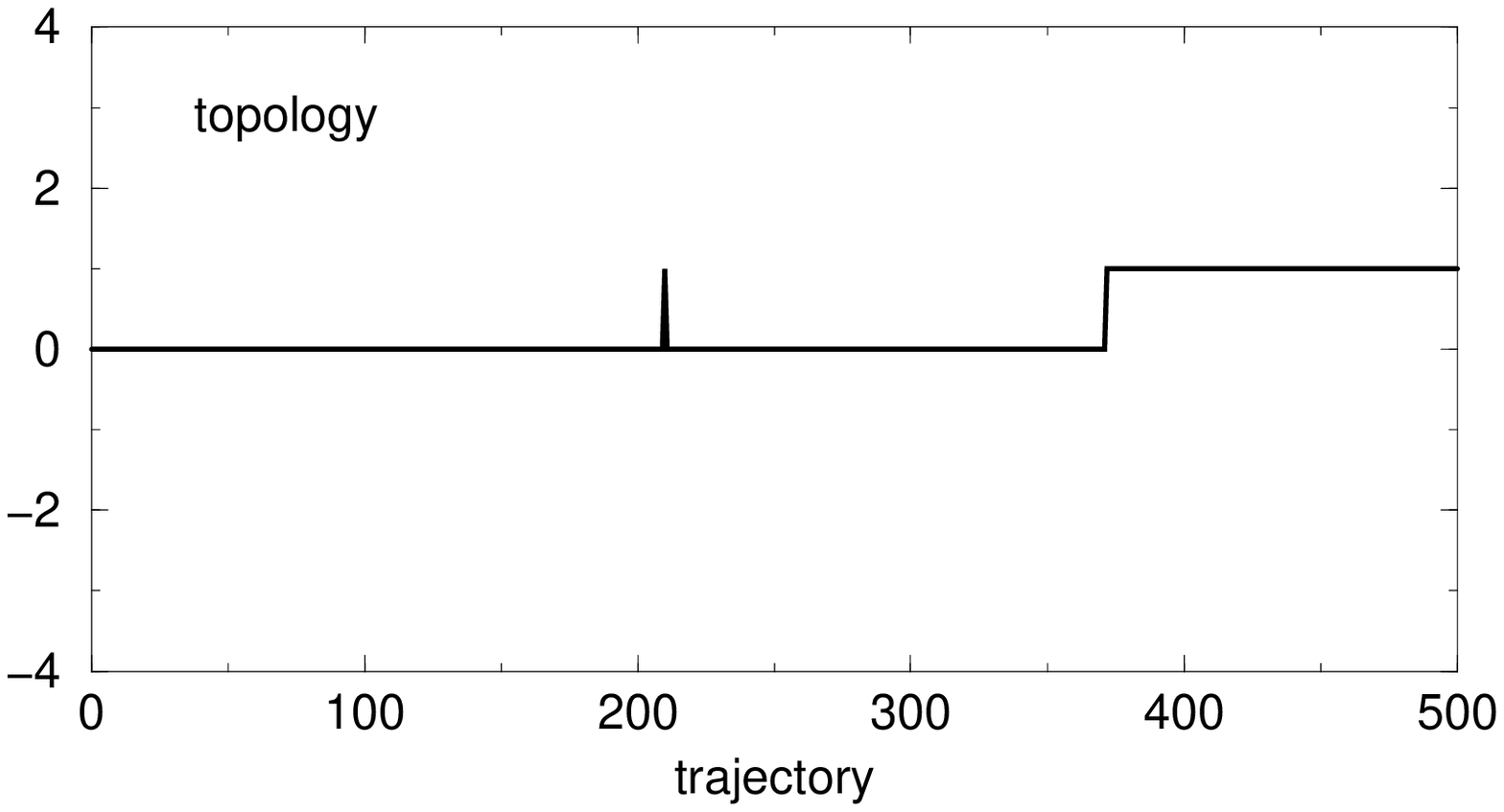}
\includegraphics[width=7.cm]{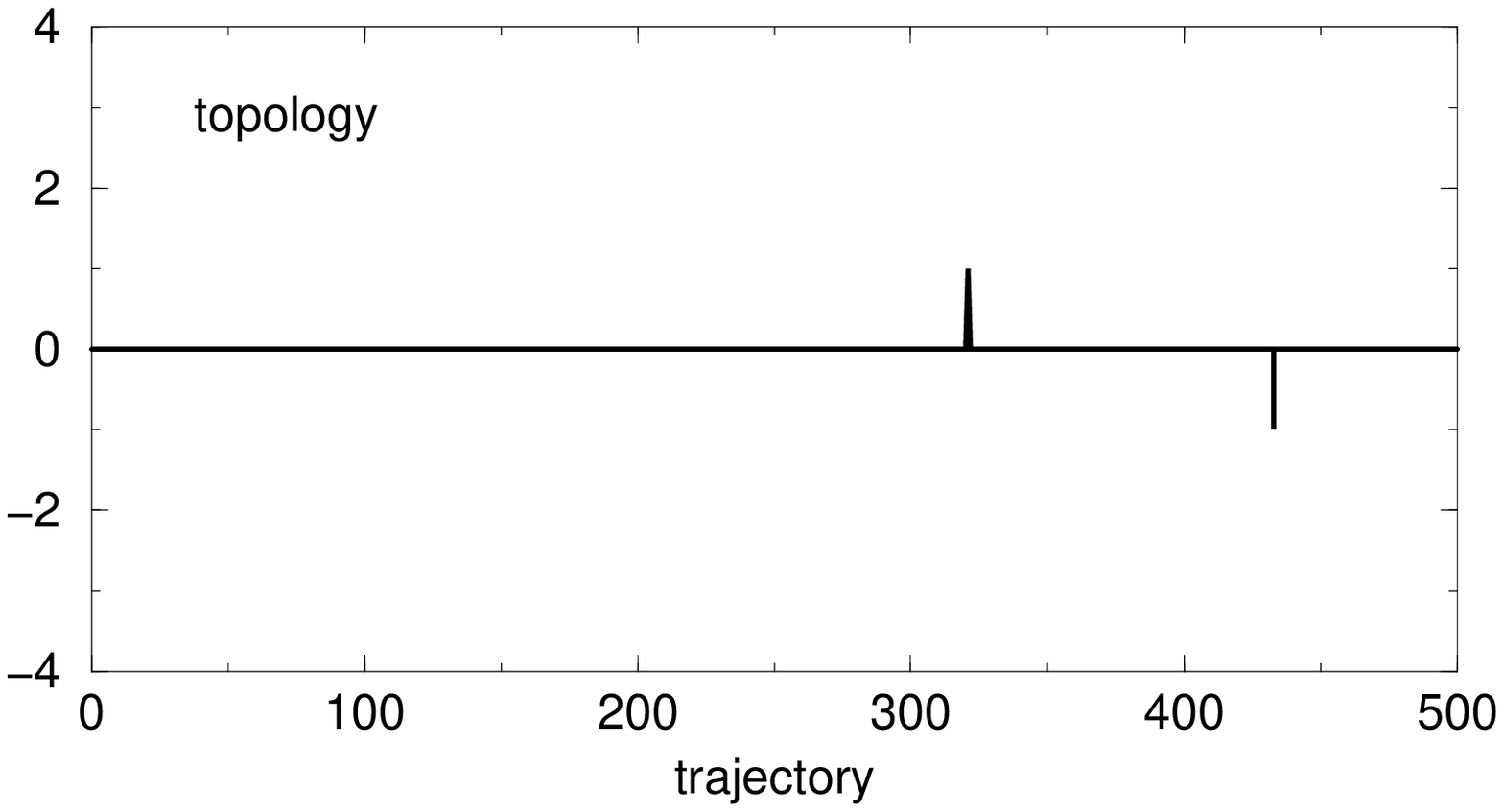}
\vspace{-0.15cm}\\
\hspace*{-0.5cm}
\includegraphics[width=7.cm]{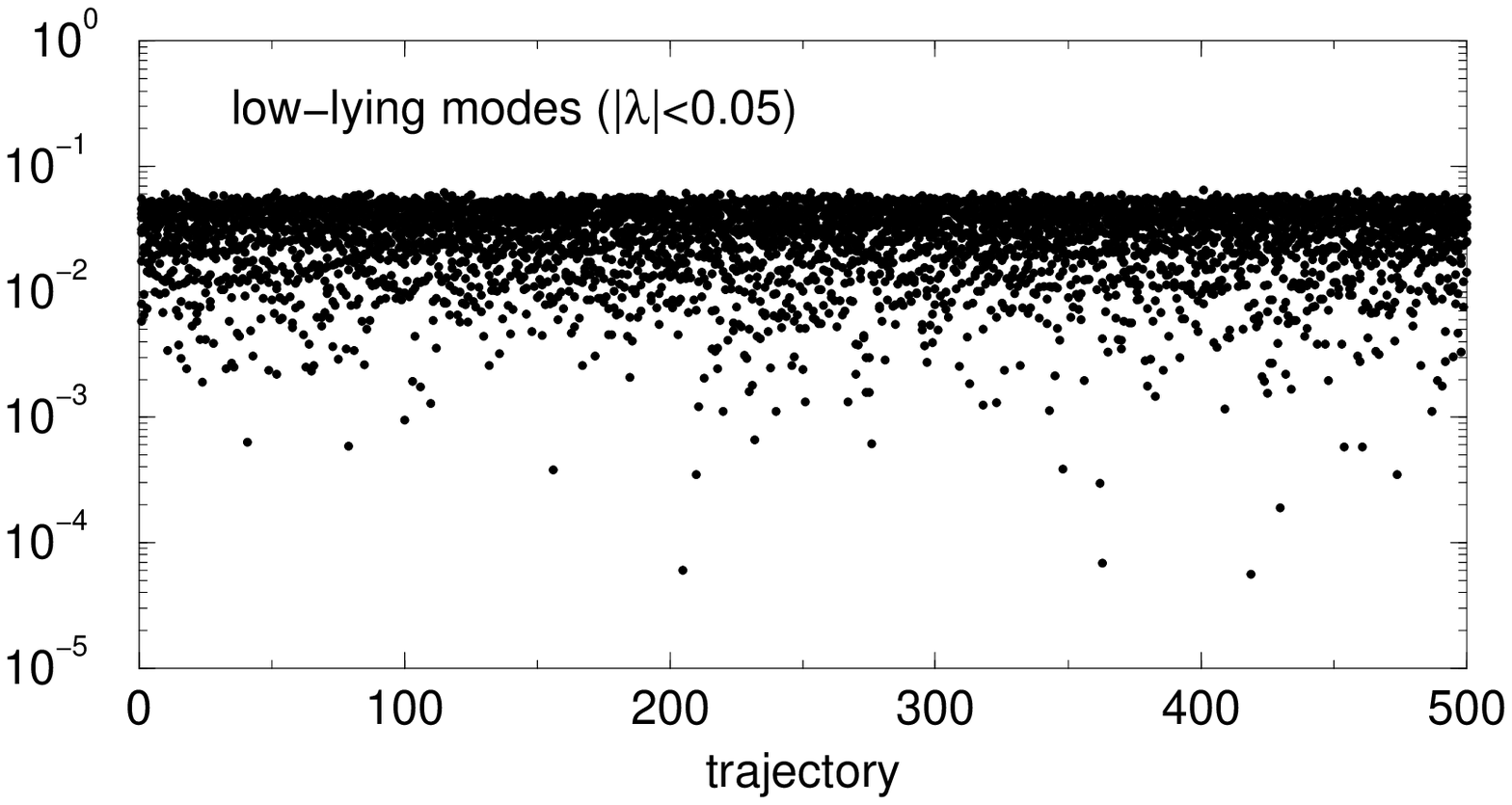}
\includegraphics[width=7.cm]{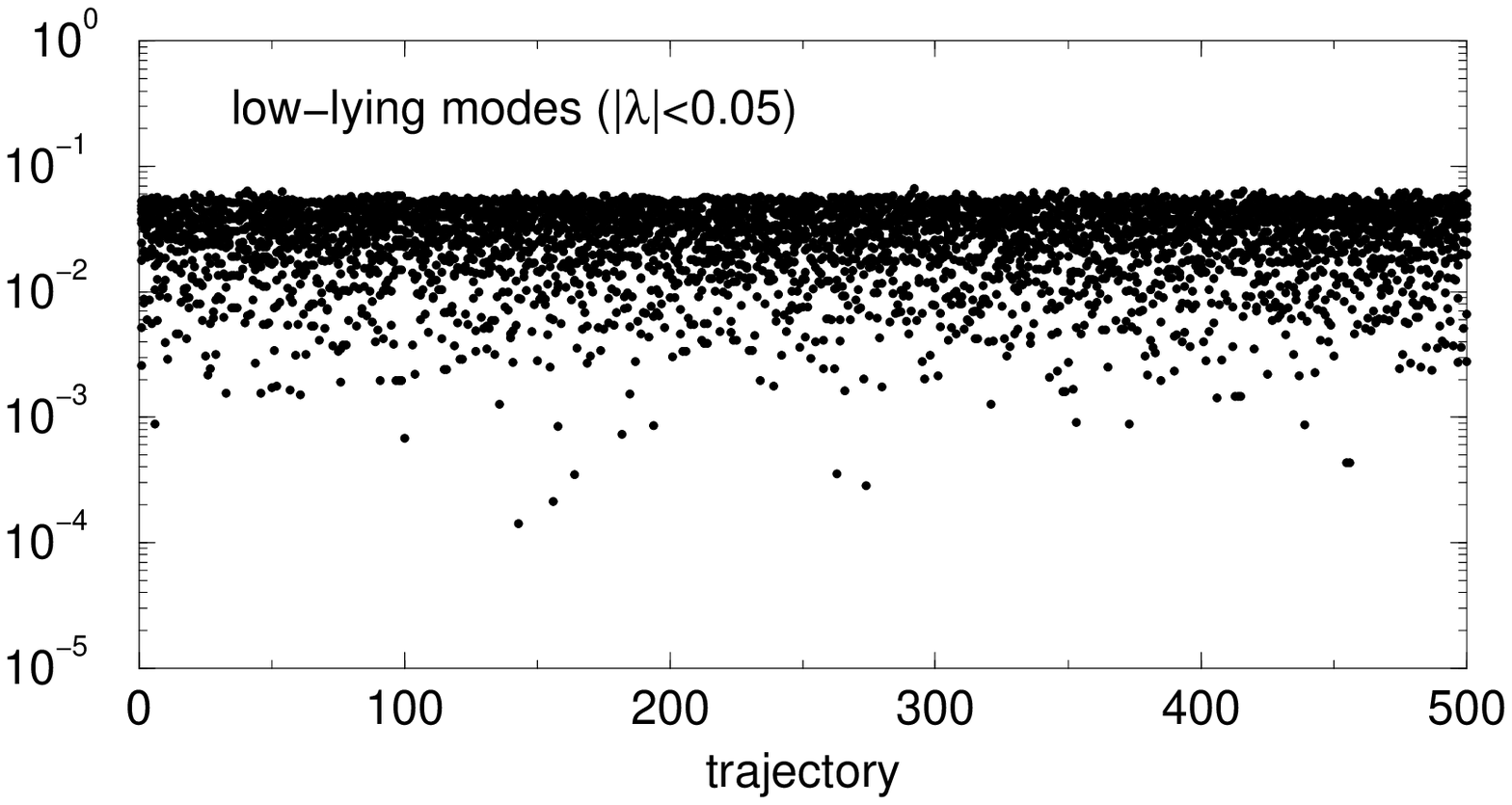}}
\vspace{-0.2cm}
\caption{
  History of plaquette (top), topological charge (middle),
  and the low-lying eigenvalues (bottom).
  Results for the plaquette gauge action ($1/\epsilon =0$) at
  $\beta=5.40$ (left panel) and $\beta=5.50$(right).}
\label{fig4}
\vspace{-0mm}
\end{figure}

\begin{figure}[tb]
\center{
\hspace*{-0.5cm}
\includegraphics[width=7.cm]{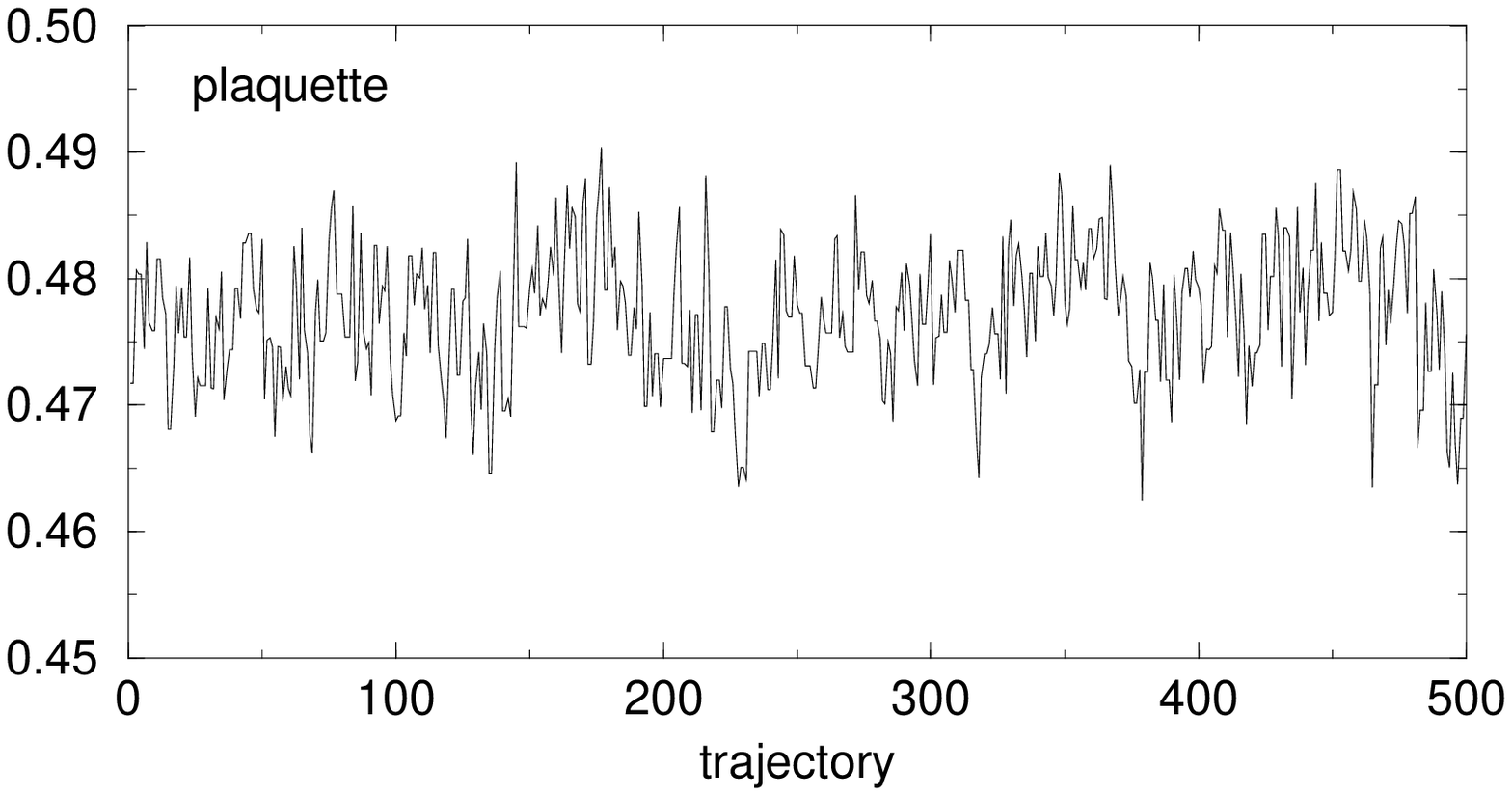}
\includegraphics[width=7.cm]{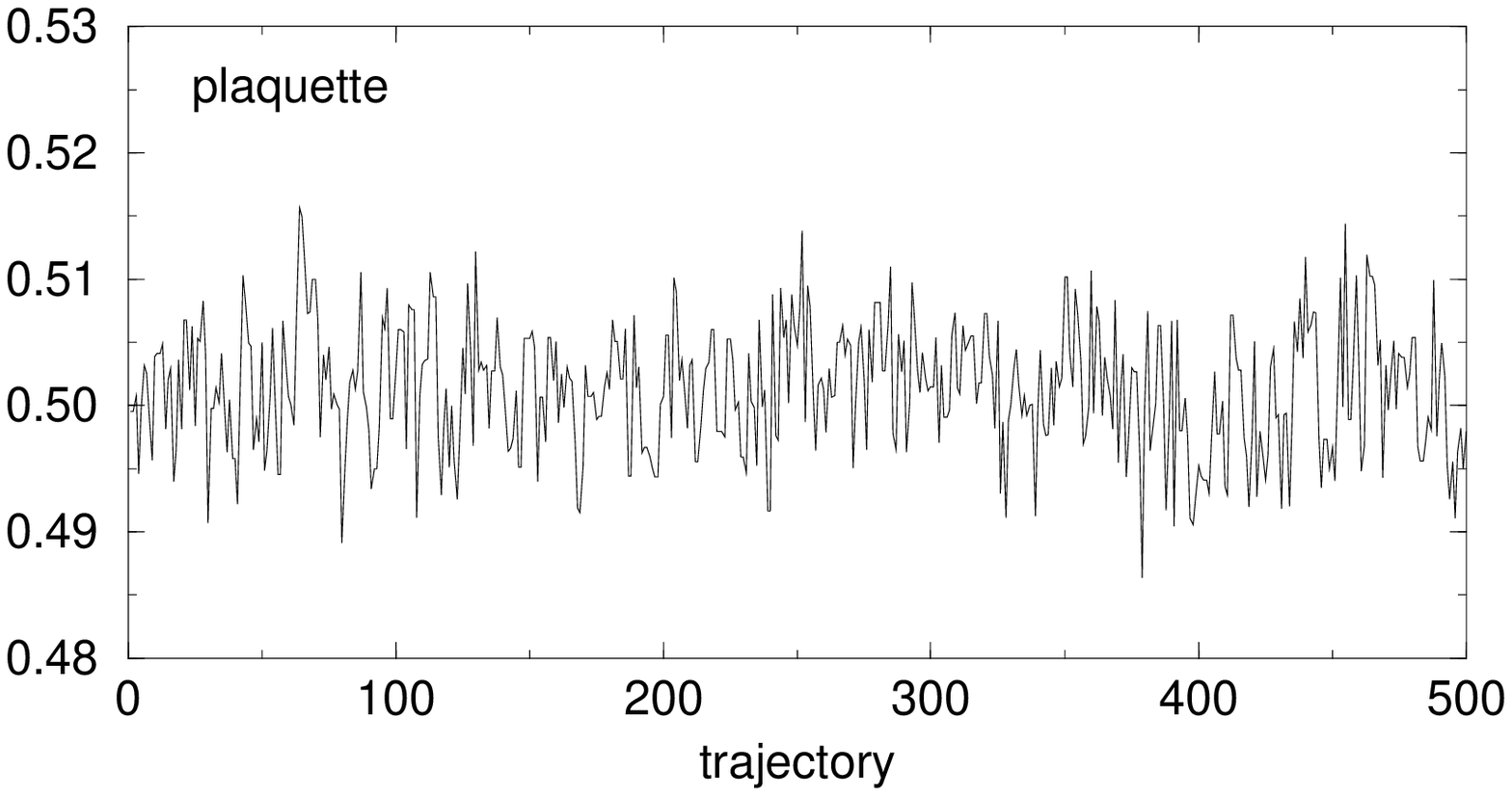}
\vspace{-0.15cm}\\
\hspace*{-0.5cm}
\includegraphics[width=7.cm]{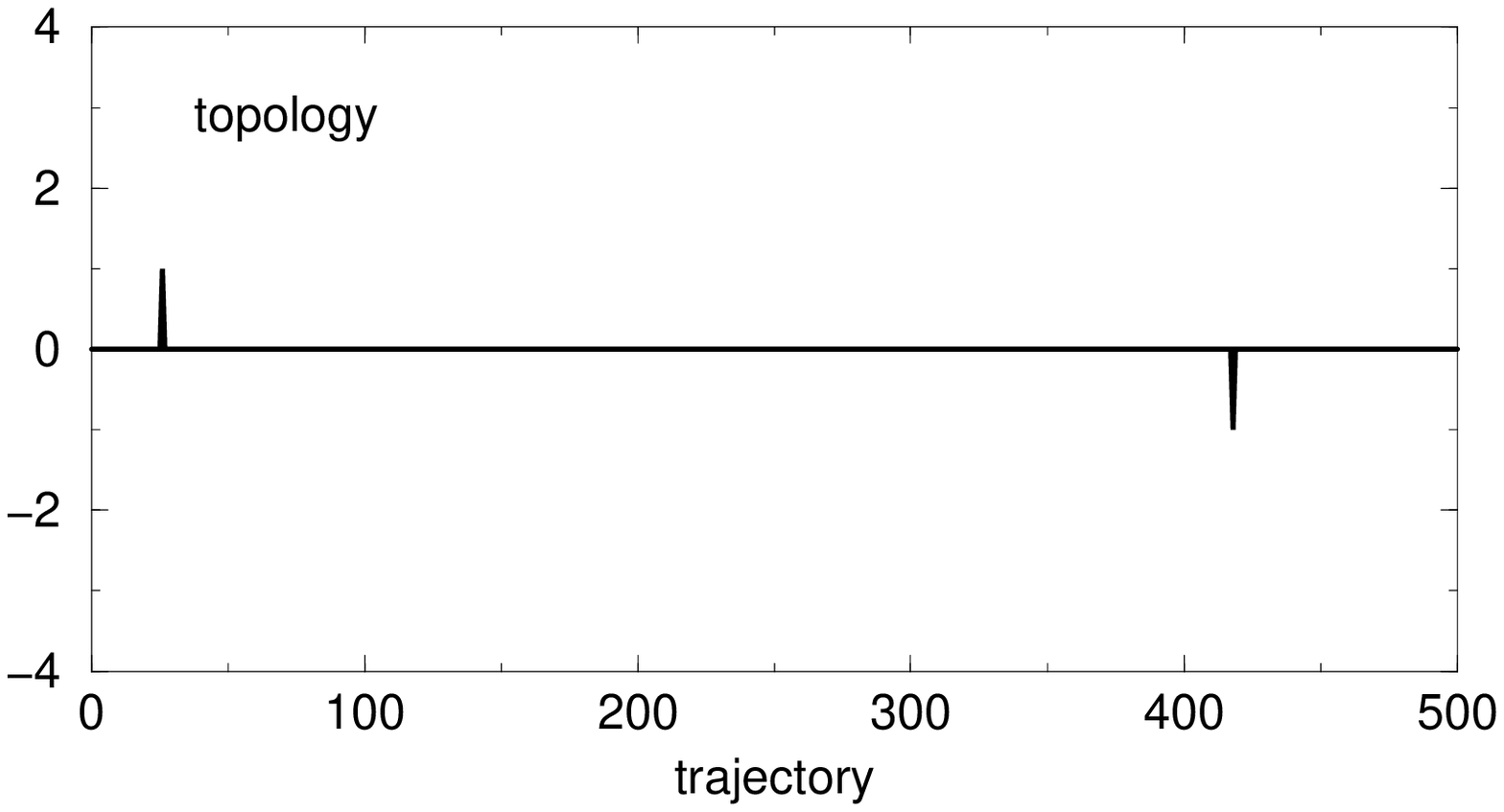}
\includegraphics[width=7.cm]{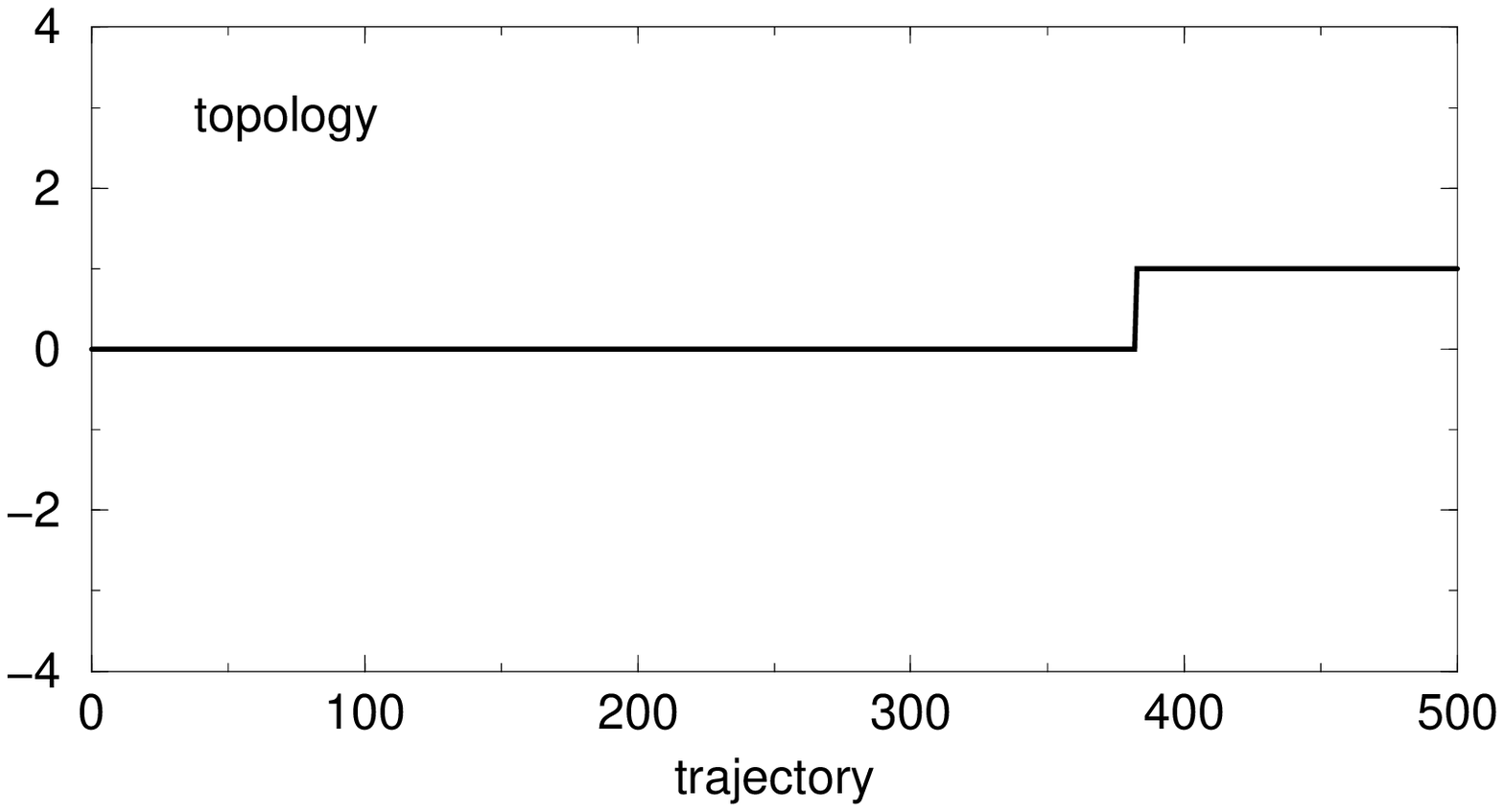}
\vspace{-0.15cm}\\
\hspace*{-0.5cm}
\includegraphics[width=7.cm]{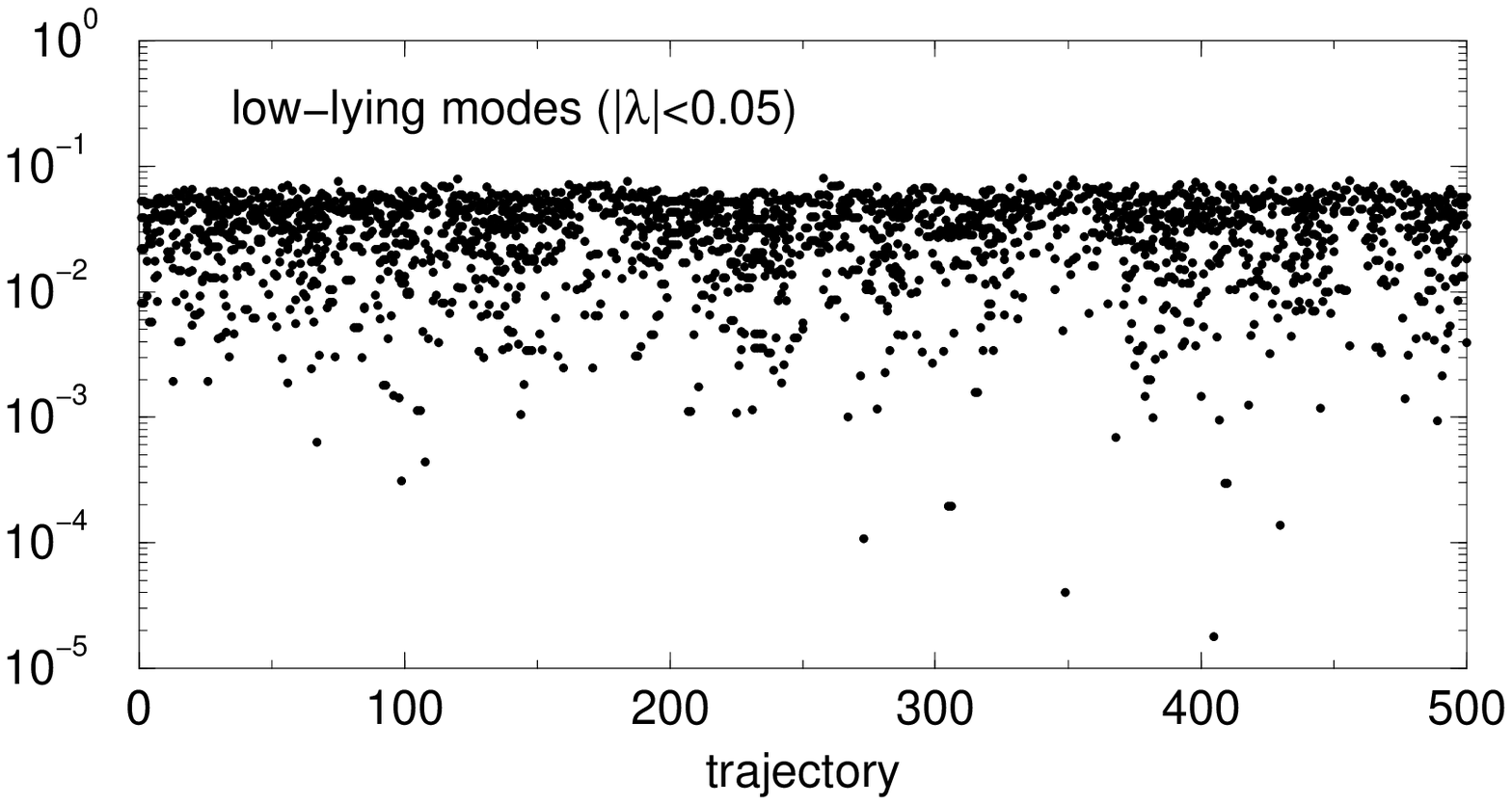}
\includegraphics[width=7.cm]{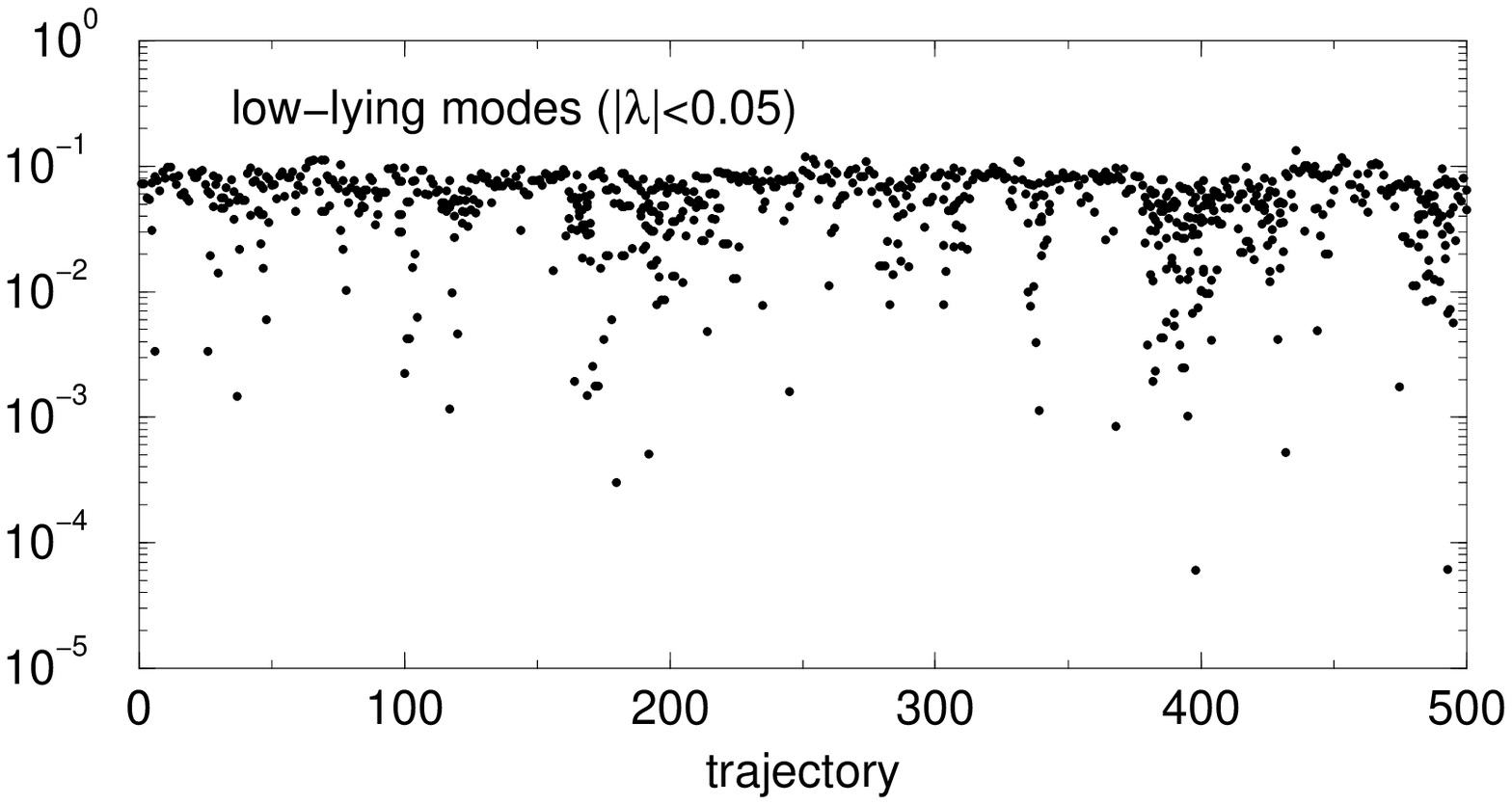}}
\vspace{-0.2cm}
\caption{
  Same as Figure~\protect\ref{fig4}, but for
  $1/\epsilon=1$:
  $\beta=0.70$ (left panel) and $\beta=0.80$ (right).
}
\label{fig5}
\vspace{-0mm}
\end{figure}

The simulation time history of the plaquette, the
topological charge, and the low-lying eigenvalues of $H_W$
is displayed in Figures~\ref{fig4} and \ref{fig5} for
$1/\epsilon=0$ and 1, respectively.
The topological charge is observed by monitoring zero-crossing
of the lowest eigenvalue of $H_W$.
No clear difference is found for the topology change
between the plaquette and the topology conserving gauge actions.
For the distribution of the low-lying eigenmodes, the topology
conserving gauge action exhibit smaller density than the
plaquette action. 

To draw definite conclusions, we of course need to compare
these actions on fixed lattice spacings on larger lattices.
The present runs are probably done on rather coarse
lattices. 
Whether the topology conserving action is practically
feasible or not will be examined in more detail.

\smallskip

This work is partly supported by the Large Scale Simulation
Program No.~136 (FY2005) of 
High Energy Accelerator Research Organization (KEK).
H.M. is supported by Grant-in-Aid of the Ministry of Education
(No. 16740156).


\begin{thebibliography}{99}

\bibitem{Neuberger:1997fp}
  H.~Neuberger,
  \emph{Exactly massless quarks on the lattice},
  \emph{Phys. Lett. B} {\bf 417} (1998) 141,
  [hep-lat/9707022].
  %%CITATION = HEP-LAT 9707022;%%

\bibitem{Hernandez:1998et}
  P.~Hernandez, K.~Jansen and M.~Luscher,
  \emph{Locality properties of Neuberger's lattice Dirac operator},
  \emph{Nucl. Phys.} {\bf B552} (1999) 363,
  [hep-lat/9808010].
  %%CITATION = HEP-LAT 9808010;%%

\bibitem{Edwards:1998sh}
  R.~G.~Edwards, U.~M.~Heller and R.~Narayanan,
  \emph{Spectral flow, chiral condensate and topology in lattice QCD},
  \emph{Nucl.\ Phys.} {\bf B535} (1998) 403,
  [hep-lat/9802016].
  %%CITATION = HEP-LAT 9802016;%%

\bibitem{Golterman:2003qe}
  M.~Golterman and Y.~Shamir,
  \emph{Localization in lattice QCD},
  \emph{Phys. Rev. D} {\bf 68} (2003) 074501,
  [hep-lat/0306002].
  %%CITATION = HEP-LAT 0306002;%%

\bibitem{Golterman:2004cy}
  M.~Golterman, Y.~Shamir and B.~Svetitsky,
  \emph{Mobility edge in lattice QCD},
  \emph{Phys. Rev. D} {\bf 71} (2005) 071502,
  [hep-lat/0407021].
  %%CITATION = HEP-LAT 0407021;%%

\bibitem{Golterman:2005fe}
  M.~Golterman, Y.~Shamir and B.~Svetitsky,
  \emph{Localization properties of lattice fermions with plaquette and
  improved gauge actions},
  \emph{Phys.\ Rev.\ D} {\bf 72} (2005) 034501,
  [hep-lat/0503037].
  %%CITATION = HEP-LAT 0503037;%%

\bibitem{Luscher:1998du}
  M.~Luscher,
  \emph{Abelian chiral gauge theories on the lattice with exact gauge
       invariance},
  \emph{Nucl.\ Phys.} {\bf B549} (1999) 295,
  [hep-lat/9811032].
  %%CITATION = HEP-LAT 9811032;%%

\bibitem{Fodor:2003bh}
  Z.~Fodor, S.~D.~Katz and K.~K.~Szabo,
  \emph{Dynamical overlap fermions, results with hybrid Monte-Carlo
  algorithm},
  \emph{JHEP} {\bf 0408} (2004) 003,
  [hep-lat/0311010].
  %%CITATION = HEP-LAT 0311010;%%

\bibitem{Neuberger:1999pz}
  H.~Neuberger,
  \emph{Bounds on the Wilson Dirac operator},
  \emph{Phys. Rev. D} {\bf 61} (2000) 085015,
  [hep-lat/9911004].
  %%CITATION = HEP-LAT 9911004;%%

\bibitem{Fukaya:2003ph}
  H.~Fukaya and T.~Onogi,
  \emph{Lattice study of the massive Schwinger model with Theta term under
  Luescher's `admissibility' condition},
  \emph{Phys. Rev. D} {\bf 68} (2003) 074503,
  [hep-lat/0305004].
  %%CITATION = HEP-LAT 0305004;%%

\bibitem{Shcheredin:2004xa}
  S.~Shcheredin, W.~Bietenholz, K.~Jansen, K.~I.~Nagai, S.~Necco and
  L.~Scorzato,
  \emph{Testing a topology conserving gauge action in QCD},
  hep-lat/0409073.
  %%CITATION = HEP-LAT 0409073;%%

\bibitem{Bietenholz:2004mq}
  W.~Bietenholz, K.~Jansen, K.~I.~Nagai, S.~Necco, L.~Scorzato and
  S.~Shcheredin [XLF Collaboration],
  \emph{Lattice gauge actions for fixed topology},
  \emph{AIP Conf. Proc.}  {\bf 756} (2005) 248,
  [hep-lat/0412017].
  %%CITATION = HEP-LAT 0412017;%%

\bibitem{Onogi}
  H. Fukaya, T. Onogi, S. Hashimoto, T. Hirohashi, and K. Ogawa,
 \emph{Parameter dependence of the topology change and the scaling 
        properties of the topology conserving gauge action},
 \emph{PoS(LAT2005)317}.
%%CITATION = HEP-LAT 0509184;%%

\bibitem{Eshof02}
%\bibitem{vandenEshof:2002ms}
  J.~van den Eshof, A.~Frommer, T.~Lippert, K.~Schilling and
  H.~A.~van~der~Vorst,
\emph{Numerical methods for the QCD overlap operator. I: Sign-function and
  error bounds},
 \emph{Comput. Phys. Commun.} {\bf 146} (2002) 203,
  [hep-lat/0202025].
%%CITATION = HEP-LAT 0202025;%%




\end{thebibliography}
\end{document}